\documentclass[11pt,twoside]{article}
\usepackage{asp2010}
\usepackage{enumerate}
\usepackage{graphicx}

\resetcounters

\begin{document}

\title{Common envelope: the progress and the pitfalls}
\author{N.~Ivanova$^1$
\affil{$^1$University of Alberta, Edmonton, AB T6G 2G7, Canada}}

\begin{abstract}
The common envelope event is one of the most important and uncertain evolutionary stages
that lead to formation of compact binaries.
While the problem is almost 30 years old, its theoretical foundation did not progress much 
from the first proposed consideration.
For many years, the simple estimate  provided by $\alpha\lambda$-formalism has been  
intensively used by population synthesis studies and, not surprisingly, frequently 
contradicted observations.
In recent years, the advancements in our studies of stellar structure, progress of the numerical techniques 
for hydrodynamical simulations as well as increase of the computer power and new observations started 
to bring improvements to our understanding of the common envelope phase.
We review main physical processes taking place during the common envelope phase
from the theoretical point of view and how they affect the values of classical formal parameters.
In particular, we discuss the energy budget problem 
-- what are the energy sources, sinks
and  what is the condition for the envelope to disperse,
as well as the importance of choosing the definition of the remnant core to the common envelope outcome.
\end{abstract}

\section{Introduction}

A key event in the formation and evolution of most compact binaries is the 
short-lived common envelope (CE) phase \citep{Os76,Pa76}. 
The CE phase could start when the largest of the two components (the donor star), or both components at the same time, 
had initiated dynamically unstable mass transfer to its companion star (see also \S2). 
This occurs, e.g., 
if the donor's envelope expands faster than its Roche lobe, and eventually engulfs the companion star.
The drag imposed by the CE transports both energy and angular momentum out of the orbit and thus shrinks the binary.

Depending on whether the envelope can be dispersed sufficiently fast, the system will end up either as a close binary or a merged single star.
The standard discriminator test is  known as the {\bf energy formalism} \citep{Webbink84,Livio88}, in which the energy difference between 
the orbital energies before and after the event is compared with the energy required to evacuate the envelope to infinity. 
The latter, denoted $E_{\rm bind}$ in the equation below, is commonly taken to be the sum of the potential energy of the envelope and its internal energy.

\begin{equation}
E_{\rm bind} = \Delta E_{\rm orb} = E_{\rm orb,i} - E_{\rm orb,f} = -\frac{ G m_1 m_2} {2 a_{\rm i}} + \frac{ G m_{1\rm,c} m_2} {2 a_{\rm f}} 
\end{equation}
Here  $a_{\rm i}$ and $ a_{\rm f}$ are the initial and final binary separations, $m_1$ and $m_2$ are the initial star masses
and $ m_{1\rm,c}$ is the final mass of the star that lost its envelope.

\begin{equation}
E_{\rm bind} = - \int_{\rm core}^{\rm surface} \left ( \Psi(m) + \epsilon (m) \right) dm  
\end{equation}
Here $\Psi(m)=-Gm/r$ is the  gravitational potential and $\epsilon$ is the specific internal energy.

For simplicity of use,  specifically for population synthesis simulations, a simplified parametrization was introduced. 
The binding energy is expressed then in terms of the donor envelope central concentration, using  a parameter $\lambda$:
\begin{equation}
E_{\rm bind} = \frac {G m_1 m_{1,\rm e}} {\lambda R_1} 
\label{lambda}
\end{equation}
Here $m_{1,\rm e}=m_1 - m_{1\rm,c}$ is the mass of the removed giant envelope and $R_1$ is the radius of the giant star at the onset of CE.

To measure the energy transfer efficiency from the orbital energy 
into the envelope expansion, another parameter, $\alpha$, was introduced. The balance of energy is written then as

\begin{equation}
\alpha {\lambda} \left ( \frac{ G m_{1\rm,c} m_2} {2 a_{\rm f}} -\frac{ G m_1 m_2} {2 a_{\rm i}} \right ) =
\frac {G m_1 m_{1,\rm e}} {R_1}
\label{allam}
\end{equation}
Since for many  stars (at least low-mass star, as detailed models show) $\lambda=1$ and $\alpha\la 1$ due to energy conservation, it is very common, if detailed stellar models are not available, to accept $\alpha \lambda = 1$. 

Another approach is to consider, instead of the energy balance, the angular momentum \citep{Nelemans00, Nelemans05}.
It is known as {\bf $\gamma$-formalism}:

\begin{equation}
\frac{\Delta J_{\rm lost}}{J_{\rm i}}= \frac{J_{\rm i} - J_{\rm f}}{J_{\rm i}} = \gamma \frac{m_{1,\rm e}} {m_1+m_2} 
\label{gammaform}
\end{equation}
Here $J_{\rm i}$ and $J_{\rm f}$ are the angular momenta of the initial and the final binaries.

$\gamma$-formalism could be used to explain the formation of 
the systems for which it was originally calibrated with observations (double white dwarf binaries).
However, it has been shown that this reverse engineering produces a range of values of the parameter 
$\gamma$ which can   not be extended to the case of the CE in an arbitrary binary  (see \citealt{Web08} and \citealt{woods10},  for  more  formal  mathematical  explanation  see \citealt{woods11_proc}). 
It has to be appreciated as well that even in the case of double white dwarf binaries 
 $\gamma$-formalism implies an energy {\it generation} during the first episode of the mass transfer,
reconstructing as such a non-dynamical event but not a CE \citep[likely, a stable mass transfer, as was shown in][]{woods11}.
More recent comparisons of the observed populations of post-CE binaries with the population synthesis models 
had also shown severe inconsistency with $\gamma$-formalism \citep{Dav10,Zor10}.
Considering recent theoretical and observational disagreements with the $\gamma$-formalism statements,  
this formalism, in its current form, can not be recommended for studies of arbitrary CE events.

Let us review in more detail the energy formalism.
There, the simple parametrization with $\alpha \lambda = 1$  had also 
started to fail once more statistics from observations came.
In particular, to explain the observed formation rate of low-mass X-ray binaries, 
1 per mln yr per Milky Way  \citep{Rom98}, it is apparently necessary to adopt
unrealistically high values of $\alpha$ \citep{Kal99}; 
e.g., \cite{Yun06} showed that the observations could be matched with  $\alpha\lambda \ga 2$.
The situation became even more dramatic with the use of values of $\lambda$ taken from detailed stellar models: 
 in massive giants $\lambda \ll 0.1$  \citep{Tau01, podsi03}. It has been shown as well that   
with  $\alpha\le 1$ only an intermediate-mass companion could  avoid 
a merger; this challenges the formation of a low-mass X-ray binary with a low-mass companion in general \citep{Justham06}.

The situation may seem dire, but progress in several subproblems has been made.
In order to understand the roots and appreciate the profundity of the problem once
the straw-clutching of parametrization -- which is done by forcefully fitting the population synthesis runs with observations -- are discarded,
one has to resolve two seemingly obvious questions:

\begin{itemize}
\item What is really $\lambda$?
\item What is really $\alpha$? 
\end{itemize}

\section{Phases of the common envelope event}

\begin{figure}
\begin{center}
\includegraphics[scale=0.6]{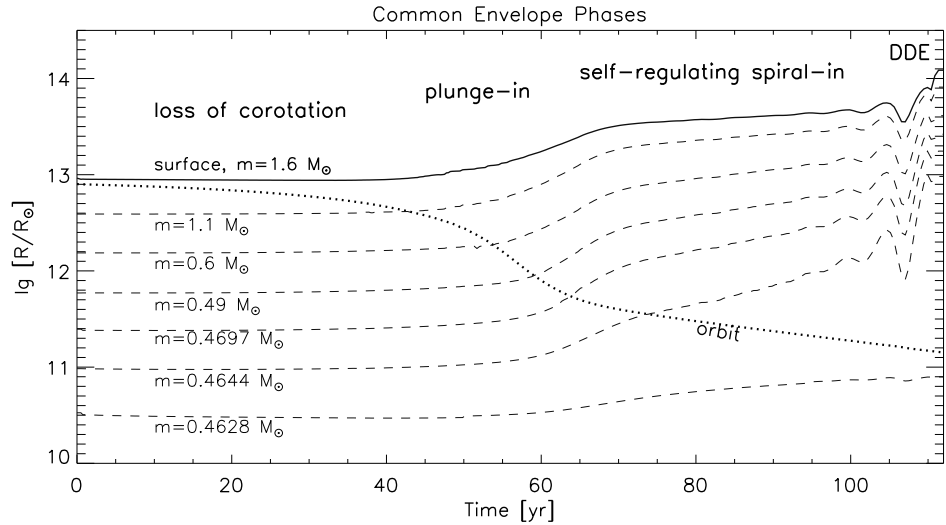}
\caption{Main phases of the common envelope. Showed the case of a 1.6 $M_\odot$ red giant and a 0.3$M_\odot$ WD. 
Dashed lines show the lines of constant mass. Used data from 1d hydrodynamical simulations in \cite{Iva02}}
\end{center}
\end{figure}

For further understanding, it is important to realise that a CE event is not a single self-similar and purely dynamical event 
for all the binaries, but most likely a sequence of 
several phases, where each operates on its own timescale \citep[][see also Fig.~1]{podsi01}:

\begin{itemize}
\item {\bf Loss of corotation}

The start of the spiral-in could be caused by, e.g.: 

{
\begin{enumerate}[(a)]
\item {a dynamically unstable (runaway) mass transfer \citep[for a linear stability analysis of the mass transfer see, e.g., ][]{Hje87, Sob97, Rit99}}
\item {a dynamical instability \citep[the Darwin's instability,][]{dar79}}, or a secular tidal instability \citep{Hut80, Lai93,Egg01},  when the spin angular momentum of the system is more than a third of its orbital angular momentum
\item {in some special cases, e.g. in nova systems because of the expansion of the nova shell engulfing the companion, or in systems with a compact companion where a stable but very fast mass transfer leads to a  formation of the extended envelope around a compact object:  e.g., reincarnation of a white dwarf in the form of a red giant \citep{Nom79, Nom07}, or ``trapping'' the accreting 
material in the form of the envelope around a neutron star in case of supereddington mass transfer \citep{Beg79, Hou91, Kin99} }
\end{enumerate}
}

The loss of the corotation occurs on a dynamical timescale, although the stellar structure at that moment is strongly affected by mass-transfer history before the dynamical instability sets on, and this stage could last hundreds of years \citep{podsi02}.

\item {\bf Plunge-in}

A rapid spiral-in, during which all the frictional energy is deposited in the envelope, drives its expansion and {\it may} lead to its dynamical ejection. This stage is purely dynamical. 

\item {\bf Slow spiral-in}

When the orbital decay transferred enough energy to cause significant envelope expansion, the spiral-in slows down. 
This is a self-regulating phase when all released frictional luminosity is transported to the surface where it is radiated away \citep{mmh79}.
This phase is non-dynamical and operates on the thermal time-scale of the envelope.

\item {\bf Termination}

Slow spiral-in ends with either the ejection of the envelope (e.g., via delayed dynamical ejection, \citealt{Iva02}, \citealt{han02}), or by a (slow) merger. 

\end{itemize}

Further evolution of the survived binary could also be affected by a circumbinary disk, winds from the giant's remnant and so on.

The dynamical plunge-in phase, when considered as a separate ``encapsulated'' event, is the best studied phase of the CE to 
date \citep[for a thorough review see ][and references therein]{Taam00,Taam10}. This stage proceeds dynamically and no 
additional energy sources or sinks affect its evolution.  
The complications arise from: 
(i) this stage usually does not end with a dynamical ejection \citep{Ric08, Taam10}, but instead hydrodynamical simulations show 
the transition into a slow spiral-in; 
(ii) the initial conditions are not properly modeled and this may affect the final results (see more details in \S5).

\section{Story of the  $\lambda$-parameter}

$\lambda$ is the factor that links the ``properly'' calculated binding energy from detailed stellar models and its parametrized form. 
To be exact, by definition it is:

\begin{equation}
- \frac {G m_1 m_{1,\rm e}} {\lambda R_1} 
		= \int_{\rm core}^{\rm surface} \left ( \Psi(m) + \epsilon (m) \right) dm 
		= \Omega_{\rm env} + U_{\rm env} = W_{\rm env}
\label{lamw}
\end{equation}
Here $\Omega_{\rm env}$ is the potential energy, $U_{\rm env}$ the internal energy and $W_{\rm env}$ is the total energy of the envelope.
If a code that calculates detailed stellar models is available, and {\bf if the core-envelope boundary is known exactly}, $\lambda$ can be  
calculated for any star at any evolutionary stage. 

Intuitively, $\lambda$ was meant to be a measure of central concentration of the star, which would translate into the expense side of the energy budget.

\subsection{Simplification of detailed stellar models}

Is some works, an over-simplification to obtain $\lambda$ from detailed stellar models 
is made, by connecting the potential energy and the internal energy as 

\begin{equation}
-E_{\rm bind} \equiv W_{\rm env} \equiv \Omega_{\rm env} + U_{\rm env} = \frac{1}{2} \Omega_{\rm env}
\label{lams}
\end{equation}
This, presumably, implies the use of the virial theorem.
It should be appreciated that this expression is not valid for the following reasons:

\begin{itemize}
\item Virial theorem indeed states that, for a star in hydrostatic equilibrium, $2K+\Omega=0$ (here $K$ is the total kinetic energy), however 
this is valid {\bf only for a the entire star}.
\item $K\equiv U$ {\bf only} in the very specific case of constant adiabatic exponent  
$\Gamma_3 \equiv \left ( \frac{\partial \ln T }{\partial \ln \rho} \right)_{\rm ad} = {5}/{3}$. This is not at all not the case in e.g., partial ionization layers, or in massive stars. E.g., in massive stars, $\Gamma_3 \rightarrow {4}/{3}$ and, since $K \rightarrow  U/2 $, then $E_{\rm bind} = W\rightarrow 0$, which is far from 
 $E_{\rm bind}=-\Omega_{\rm env} /2$ as in the Eq.~(\ref{lams}).
\end{itemize}

If one defines $\lambda_{\rm g}$ as only dependent on the potential energy \citep[as in][]{Tau01}, 
and so $\lambda_{\rm g}/\lambda=E_{\rm bind}/\Omega_{\rm env}$, then simplification as in the 
Eq.~(\ref{lams}) would predict
$\lambda_{\rm g}/\lambda=1/2$; however detailed stellar models, as expected, show a very different ratio (see example on Fig.~2).


\subsection{A constant value: how good is that?}

In the case when detailed stellar models are not available, the most commonly accepted value is $\lambda=1$. 
This value is rather good approximation for low-mass red giants, especially with masses $\sim 2M_\odot$. 
In a general case, however, $\lambda$ is the function of mass and radius, and it changes as the star evolves. 
For most stars, $0.1 <\lambda < 45$ \citep{Dew00}, and massive stars could also have $\lambda <0.1$ \citep{podsi03}. 

The Eq.(\ref{allam}) tells that the adoption of $\lambda=1$ leads to about the same error in the final binary separation by the same factor as a ``properly'' found $\lambda$ differs from 1.
Tabulated values of $\lambda$ for stars as a function of mass and radius, and/or fitting formulae, can be found in \cite{Slu10}, \cite{Lov10} 
and \cite{Xu10}.

\subsection{Uncertainty given by the boundary problem}

\begin{figure}
\begin{center}
\includegraphics[scale=0.6]{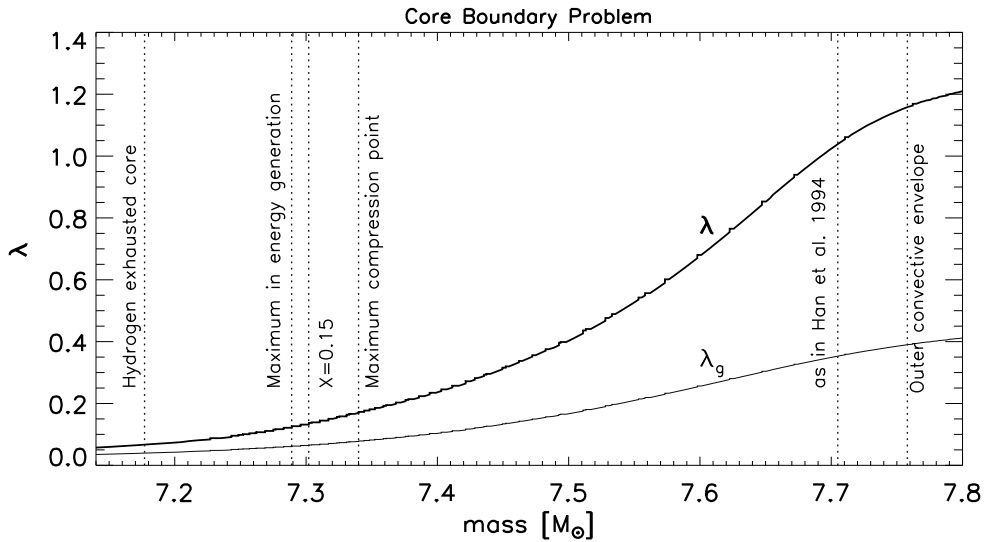}
\caption{$\lambda$ as a function of mass shown on example of 20 $M_\odot$ star when it has $R=950 R_\odot$ ($Z=0.02$, overshooting $0.2$ of the pressure scale
and no wind loss). For comparison $\lambda$ as in Eq.(2) (thick solid line) and $\lambda_{g}$  (thin solid line, see discussion \S3.1) are shown. Dotted lines correspond to several possible core definitions, as discussed in \S3.3, 3.4.}
\end{center}
\end{figure}

Even with detailed stellar models, we still need to decide on where to place the boundary between the core and the envelope.
It has been first noticed in \cite{Tau01}, that placing this boundary within the hydrogen shell between the He core and the 
bottom of the convective envelope, $\lambda$ can change by a factor of 10-70, depending on the star, where the discrepancy is more marked 
for more massive stars.  This is explained by the fact that the binding energy of the H shell far exceeds the binding energy
of the convective envelope above. 

The uncertainty in the value of $\lambda$ due to an ambiguous core definition, as discussed in \S3.2, 
translates directly to a similar uncertainty in the final binary separation. It is clear therefore that the ambiguity in core definition 
leads to the same or even larger uncertainty than the use of a constant $\lambda$ for all the stars. Worse still, this uncertainty 
is invariably overlooked in the studies of CE outcomes, and the accepted definition of the core in a particular set of simulations 
is usually the least clearly discussed condition. Note that the vagueness in placing the core boundary also comes from the fact that 
the remnant of a giant in the formed binary (immediately after the envelope ejection) does not have to be the same as the post-CE remnant 
we observe today, as  the remains of the Hydrogen shell could be later removed by strong winds, similar e.g. to Wolf–-Rayet winds.

A variety of possible definitions of where is the boundary between the remaining core and the ejected envelope could be found in literature:

\begin{itemize}
\item the minimum possible core mass is  at $X=0$ (bottom of H shell)
\item at the maximum nuclear energy generation within H shell  \citep{Tau01}
\item at the maximum nuclear energy generation plus the condition on the thickness of the remaining envelope, which itself is 
  a function of the evolutionary status of the donor  \citep[for low mass red giants and asymptotic giant branch stars,][]{Mar11}
\item where the nuclear energy generation falls below some threshold  \citep{Mar11}
\item the central mass which contains less than 10\% hydrogen \citep{Dew00}
\item core is everything below the location where $X=15\%$ \citep{Xu10}
\item where $\partial^2 \log \rho /\partial^2 m =0 $ within H shell \citep{Bis98}
\item where the function of the binding energy $y=\sinh^{-1}(E_{\rm bind})$ has the transition  
between a sharp increase and a fairly slow increase in the outer envelope \citep{Han94}
\item where the entropy profile has a transition between its increase and flat part \citep{Tau01}
\item where the effective polytropic index has discontinuity \citep{Hje87}
\item the bottom of the convective envelope (top of the H shell) gives the maximum possible core mass, as convective envelopes are known to expand upon mass loss.
\end{itemize}

Last three definitions describe the same place, the transition between the radiative zone of H shell and the convective envelope. 
Note that not all definitions are applicable to every star. 
Some are created only to low-mass giants or asymptotic giant branch stars 
(e.g., thickness of the remaining envelope), some conditions can not be either found or determined uniquely in all the stars 
(e.g., the condition on $\partial^2 \log \rho /\partial^2 m = 0 $  can not be determined in case of $20 M_\odot$ as on Fig.~2). 
Unfortunately, none of the above definitions has a solid physical foundation explaining the choice.


\subsection{Core boundary consideration via post-thermal readjustment}

Another method to define the core remaining after a CE event was proposed in \cite{Iva11}. 
The idea is that different parts of the H shell react differently to the very rapid mass loss;
both immediately after the envelope ejection and subsequently on the thermal timescale. 
Indeed, the bifurcation in the reaction of H shell material -- whether the shell re-expands after the mass loss or collapses -- has been found long ago 
for low-mass giants \citep{Dei70}. In \cite{Iva11} it was shown that similar divergence points exist in the H shell of more massive giants as well. 
If mass is removed below such a divergence point, the remnant contracts on thermal timescale from its post-CE state, if the mass is left above such 
divergence point, the star expands during its thermal readjustment; it could experience strong thermal pulses, or even develop convective envelope. 
The subtle difference with the low-mass case is that, immediately after the fast mass removal, the remnant, as it contains a non-degenerate core, 
always starts to expand compared to its pre-CE state \citep[similar expansion for donors with non-degenerate cores was found, e.g., in adiabatic 
studies by][]{del10}.

The divergence point above coincides with the location where $P/\rho$ reaches its maximum value within the H shell, 
so called ``compression point'' $m_{\rm cp}$. In general, this divergence point does not coincide with any of the ones listed in \S3.3.

Finding  the core mass via  $m_{\rm cp}$ provides an upper bound on the {\it maximum} post-thermal readjustment core mass as it is not 
possible to completely rule out that the dynamical phase does not remove mass below $m_{\rm cp}$.
Such extra mass removal however is not likely to happen if the characteristic orbital evolution time during the final stages of the spiral-in
is comparable to the core thermal response time near $m_{\rm cp}$ point -- which is as short as 10-100 years. If the event proceeds on this timescale,
then the core material below $m_{\rm cp}$ starts to contract and precludes further mass removal.

On the other hand, all the mass that could have been left above $m_{\rm cp}$ is guaranteed to be pushed out by the thermal timescale mass 
transfer during star's readjustment. 
If the envelope was ejected during a dynamical plunge-in phase and the core mass is $m_{post-d}> m_{\rm cp}$,
then the core is reduced to $m_{\rm cp}$ without any additional external energy source like orbital energy, 
but only using core's own thermal energy.

The above property of $m_{\rm cp}$ can be stated as an alternative definition for the bifurcation point: 
the energy expense required to shed the envelope down to $m_{\rm cp}$ is minimal.
It is fully reasonable to remove less of the envelope (and have a bigger post-CE mass) once
$E_{\rm bind} < \alpha E_{\rm orb}$.  The thermal readjustment phase will then remove mass down to $\sim m_{\rm cp}$.
However, energy-wise, it is not advantageous to remove the envelope deeper than $\sim m_{\rm cp}$.

\bigskip

\noindent The complete understanding what exactly is $\lambda$ 
(and in other words what is the energy required to remove the envelope) however is linked with how the envelope 
ejection happens, on which timescale and what energy can be used during the CE, and therefore is amalgamated into the
$\alpha$-parameter problem.

\section{Story of the $\alpha$-parameter}

$\alpha$ is the parameter that was introduced to relate
the available energy source (orbital energy) to the energy required to disperse the envelope, where the latter
is assumed without questioning to be the binding energy. The energy balance is then written as:

\begin{equation}
\alpha \Delta E_{\rm orb} = E_{\rm bind}
\end{equation}

\noindent Energy conservation restricts $0 < \alpha < 1$.

\subsection{What energy is really required?}

The original assumption that the energy requirement equals $E_{\rm bind}$ carries one of the two tacit assumptions:
that an envelope is dispersed once its $W_{\rm env}>0$, or that an envelope with $W_{\rm env}>0$ is unstable.
This approach has been questioned in \cite{Ivach11}, for the following reasons:

\begin{itemize}

\item a star with $W>0$ can be {\it ``kinetically stable''}
\citep[i.e. additional energy is required to overcome an energy barrier between the bound and unbound states,][]{BKZ67}. 

\item  a star or its part (e.g. envelope) is unstable against adiabatic perturbations if
the first adiabatic exponent $\Gamma_1$ is approximately a constant throughout the star \citep{Chiu68} and $\Gamma_1<4/3$.
This condition however is not the same as $W>0$, which requires a constant third adiabatic exponent $\Gamma_3< 4/3$ \citep{HKT04}.
The relation between the adiabatic exponents is $\Gamma_1=1+\Gamma_1 \nabla_{\rm ad}$, where $\nabla_{\rm ad}$ is the 
adiabatic temperature gradient.
\end{itemize}

The energy conservation equation in Lagrangian coordinates applied to each mass shell in the envelope 
is written as (see, e.g., \citealt{CW66, KS72}):

\begin{equation}
\frac{\partial}{\partial t} \left(\frac{u^2}{2} + \Psi +\epsilon\right )  +
  {P}\frac {\partial (1 / \rho) }{\partial t}  = 0 \ .
\end{equation}
Here  $u$ is the velocity, $P$ is the pressure and $\rho$ is the density. 
When  heat $Q$ (e.g., frictional energy from orbit's dissipation) is imposed into the envelope with some  
arbitrary distribution ${\delta q(m)}>0$  (${\delta q(m)}$ is per  mass unit), such that 

\begin{equation}
\int_{\rm core}^{\rm surface} \delta q(m) dm = Q\ ,
\end{equation}

\noindent the energy conservation for any given lagrangian shell can be rewritten as a version of the Bernoulli equation:

\begin{equation}
(\delta q(m) + \Psi + \epsilon + \frac{P}{\rho})_{\rm start} 
	= (\frac{1}{2} u^2 + \Psi + \epsilon  + \frac{P}{\rho})_{\rm exp}  
	= \Sigma = const
\label{bern}
\end{equation}
A Lagrangian shell then 
will reach the point of no return in its expansion when its $\Sigma>0$ \citep[e.g.,][]{BKZ67,SK72}.
Past stability analyses showed that a star starts outflowing once its envelope obtained positive $\Sigma$ \citep[][and the references therein]{BK02}.
Same will occur in the case  a \emph{part} of the common envelope has obtained positive $\Sigma$,
and the whole common envelope should fully disperse if  the entire envelope would have positive $\Sigma$:

\begin{equation}
Q + \int_{\rm core}^{\rm surface}\left (\Psi(m) +  \epsilon(m) + \frac{P(m)}{\rho(m)} \right ) dm  \ge 0 \ .
\label{bernsimple}
\end{equation}

The development of such quasi-steady surface outflows operates on the timescale 
on which the envelope redistributes the dumped heat,
i.e. the thermal timescale of the envelope, or the same timescale on which  the self-regulating slow spiral-in operates.
These outflows are not likely to take place during a dynamical plunge-in phase. It is important, that  
due to the presence of a non-negative term $P/\rho$, this condition occurs before 
the envelope's total energy become positive and as such eases the survival of a binary with a smaller by mass companions.
In particular, the use of this criteria instead of the original energy formalism allows the formation
of low-mass X-ray binaries \citep{Ivach11}.


\subsection{Other energy source}

In studies of binaries that, by the energy formalism, are doomed to merge,
it was found that the {\it nuclear} energy source could play a role in the envelope ejection 
\citep{Iva02,Iva03}. 

Such binaries enter the common envelope as any other binary, but do not merge
during their dynamical plunge-in phase, proceeding to the self-regulating spiral-in phase.
If the companion is a main sequence star,
then at the end of the slow spiral-in it
starts to overfill its Roche lobe while the core of the giant is still well within its Roche lobe.
The mass transfer is forced by continuing orbital shrinkage due to drag from the envelope,
its rate is increasing and, depending on secondary and the mass ratio, is from $0.01$ to $100~ M_\odot$ per yr.
A thick stream of hydrogen-rich material is able to penetrate deep towards the core, 
reaching even the He burning shell \citep{ips02}.
The outcome of the steam penetration depends on the initial stream entropy. The latter depends on the mass
of the secondary: a lower mass main sequence star has a lower entropy, and its material is able to be
injected deeper in the giant core.

In some cases, the outcome of such rather slowly proceeding merger 
is the complete explosion of the He shell. The released nuclear energy during explosive hydrogen
burning  could exceed the binding energy of the He shell, which is in massive stars can be about few $10^{51}$ ergs. 
The rest of the common envelope has much smaller binding energy and is ejected during the same explosion as well, leaving behind a compact 
binary consisting of the core of the giant and whatever remains from a low-mass companion after the mass transfer. 
The effective $\alpha$, compared to the available orbital energy, is up to 100.

The main consequences of this {\it explosive CE ejection}  are \citep{podsi10}: 

\begin{itemize} 
\item a less massive companions could survive the CE,  
this helps the formation of low-mass X-ray binaries and could explain their observed formation rates
\item a fast rotating, hydrogen exhausted core is formed. This core 
could be a long $\gamma$-ray burst progenitor and Ic type SN progenitor at the same time,
explaining their observational connection.
\end{itemize}

\subsection{``Useful'' energy}

One of the important questions is which fraction of the available internal  
energy can be used to eject the envelope. Arguments both for \citep{Han94, Han95, han02} and against \citep{Sok02,Sok03} using the recombination energy 
have been advanced. The question is not yet settled, even though it has very important consequences
discussed in details in \citep{Web08}: if this energy should be included, then some of common envelope events
begin with {\it positive} total energy of the giant envelope, bringing the effective value of $\lambda$-parameter 
below zero! 

We note in addition that recombination energy leads to yet another attempts of parametrization in some works:
e.g. \cite{Han95} introduces an ``efficiency of use of internal energy'' and thus special second $\alpha$-parameter, 
$\alpha_{\rm th}$. 
\cite{Web08} instead proposed to define an ejection efficiency $\beta$
that has the desired property, $0\le \beta\le 1$:

\begin{equation}
\beta \equiv \frac{\Omega_{\rm env}}{\Delta E_{\rm orb} - U_{\rm env}}.
\end{equation}

\noindent The latter definition however only serves to hide the unsightly blemish of $\alpha$ not falling within the range $0\le \alpha\le 1$.
Overall, attempts to parametrize the situation with some sort of a new parameter instead of simply writing the full energy balance could 
only confuse the reader.

\subsection{Energy losses}

The energy balance between the available and required energies  as in the Eq.(1) implies that all the heat is transferred 
into the expanding envelope and that dispersed envelope has zero velocity at infinity. This is however far from the results 
obtained in detailed simulations:

\ 

\noindent{\bf Radiative losses:}

\noindent If the CE did not end during the dynamical plunge-in phase, it proceeds with the slow self-regulating spiral-in.
In this case, {\bf most} of the orbital energy dissipates at those shorter orbits, when the envelope is already expanded (see also Fig.~1).
While the giant donor star has expanded, its evolves along the giant track toward higher luminosities and keeping almost the same effective 
temperature. Its luminosity $L$ is then increased as $\propto R^2$, 
while the nuclear energy sources do not catch up with the increased $L$. 
It is the orbital energy that is converted into internal energy and then is radiated away, as for all stars evolving on thermal timescale.
It is possible to loose this way a significant fraction of the available orbital energy. 
E.g., we find that a binary consisting of $20$ and $5$ $M_\odot$ companions is loosing during the slow spiral-in 40\% of the released orbital 
energy by radiating it away from the surface of the CE, in case of $1 M_\odot$ companion the slow spiral-in stage lasts very shortly and not 
much energy is lost, but in the case of a $1.6~M_\odot$ giant and a 0.3 $M_\odot$ companion the losses are close to the 100\% of the released 
orbital energy until the delayed dynamical ejection starts \citep[data is from the simulations described in][]{Iva02}. 

\ 

\noindent{\bf Non-zero velocity at infinity:}

\noindent It is currently adopted for simple analysis that at the infinity velocity of the expanding envelope $u=0$: 
this is how the full energy conservation equation Eq.~(\ref{bern}) is translated into 
simplified energy formalism like in the Eq.~(\ref{bernsimple}) or E.~(1). However, hydrodynamical 
simulations performed with use of a smooth particle hydrodynamics, show that the kinetic energy of the ejected envelope at the infinity 
could be from 15\% to 70\% of $|E_{\rm orb,f}|$
(data is from simulations performed for \citealt{Lombardi06,Iva10}).  Values of the kinetic energy 
do differ depending on the mass ratio of the interacting stars and also on the core to mass ratio in the interacting giant.

\bigskip

\noindent The consideration of just two possible ways to loose energy compared to the pure conversion of the available energy into 
only the envelope ejection energy shows that the {\it effective} $\alpha$ (meaning that all possible energy sources are also taken 
into account, not only the orbital energy) is not only less than 1, but also is a function of the stellar masses and mass ratios 
in the binary system. There is no reason whatsoever to expect one value of $\alpha$ to apply to all CE events.

\section{Start of the CE and its influence on $\alpha$ and $\lambda$}

A not particularly well studied CE phase is the {\it onset on the CE} (see Fig.~1).
This is not an immediate process and involves the evolution of the donor before the dynamical evolution stars.
The whole list of possible unsolved problems during the initial stage when the binary is loosing its corotation includes such questions as 
the possibility for an enhanced wind losses just prior to the Roche lobe overflow, dependence of the critical mass ratio, as a condition 
to start runaway mass transfer, from the giant structure. Even without consideration of this, we know that hydrodynamical simulations 
experience the problem to start the CE. One of the important factors is how far the donor is out of corotation with the binary.

In order to have successful start of the CE, 
the initial conditions  for common-envelope hydrodynamical simulations are such 
that the dynamical plunge-in phase is literally forced to start quickly. 
It can be  done by either considering a donor  being, e.g., not in a corotation with a binary \citep[as in ][]{Ric08}, 
or by having some degree of the Roche lobe overflow initially, or by introducing initial tidal perturbations (some variation of non-corotation).
Choosing the degree of non-corotation leads to different final results for the ejected masses and the masses of formed 
circumbinary envelopes.
A fully self-consistent solution in term of initial condition 
so far is obtained only in case of physical collisions between
a giant and an intruder, where initial non-corotation is a physical sound condition \citep[e.g.,][]{Lombardi06}. 
However these simulations are always accompanied by
a significant kinetic energy of the ejected envelope at infinity and it is not clear that they can be extrapolated 
to the case of a normal CE.
Nonetheless, it is clear that the choice of the starting condition for the dynamical plunge-in phase
affects the final $\alpha$ by taking away different amount of energy in the form of the kinetic energy at the infinity.

Another reason for either $\lambda$, or the combination of $\alpha \lambda$ to be affected by this not well studied phase
is that the {\it dynamically unstable} mass loss/mass transfer in fact may start not immediately. It could have been preceded by either 
stronger than usual winds \citep[just prior the full Roche lobe overflow, e.g.][]{Egg02}, 
or the mass transfer become dynamically  unstable only after a significant fraction of the envelope has been transferred 
on thermal timescale \citep{woodsmt11}, or, if mass transfer is super-Eddington, the accretion energy released 
in the rapid mass-transfer phase unbinds a significant fraction of the giant's envelope \citep{Beer07}. 
Effectively, in all three cases, the envelope mass is decreased.
This reduces how much energy has to be spend then later on to expel the formed common envelope.
It is not entirely clear with which of the two parameters this decrease in mass should be associated.

\section{Conclusions}

As of today, we can list the main pitfalls in studies of the CE using parametrization.  Those pitfalls should be either 
avoided, or carefully discussed in terms of the (large) uncertainties they provide for the population of post-CE binaries:

\begin{itemize}
\item in general, {\bf no parametrization} can be fully trusted 
\item the parametrization in the form of $\gamma$-formalism can be used only to replicate double white dwarf binaries formation, 
similar to those used for calibration
\item the parametrization for $\lambda$ can not use a constant value: $\lambda$ always depends on mass, radius and evolutionary state 
of the star, and $\lambda$ value can differ from 1 by up two order of the magnitude.
\item value of $\lambda$ depends as well on the choice of the remnant core mass,
and with different core masses your mileage may vary as well by up to two orders of the magnitude
\item Energy sources such as recombination energy can lead to a {\it negative} value of $\lambda$
\item a deposition into the common envelope of the energy equal to $E_{\rm bind}$ does { not}
guarantee the CE ejection!
\item an assumption $0\le\alpha\le1$ is not valid in general case: $\alpha$ { can be} negative if recombination energy should be included, 
can be a factor of few if the envelope is removed by adiabatic outflows, and also can be by up to 100 if nuclear driven
ejection occurred.
\end{itemize}

As we  shun the parametrization approach and deal with understanding 
the underlying {\it physics} of the CE event, a much clearer picture emerges:

\begin{itemize}
\item it is likely that we understand now what exactly is the post-CE core, however, its mass is 
not parametrizable in an easy way
\item there is progress in understanding of the ejection condition during the slow spiral in.
A self-regulated spiral-in results however not only in another criterion to
eject the envelope, but also in more energy losses, giving as a result $\alpha$ that can vary from 0.4 to 3 
\item certain binaries could have a very large effective value of $\alpha$, up to 100, 
if they e.g. experienced an explosive common envelope event
\item recent hydro simulations showed that we are almost able to model a complete common envelope event 
if that includes only dynamical plunge-in phase, though a proper onset on the CE 
(which is an {\it initial condition} for the dynamical phase), as well as the modeling of 
self-regulating spiral-in stage is still a challenge
\end{itemize}

I cannot stress enough that there is no reason to expect either $\lambda$ or  $\alpha$ to 
have unique values for all the binaries, as they do depend 
on the evolutionary status of the donor, the mass ratio, the initial orbital separation -- 
in short, on every parameter and physical process that is involved in setting up a CE.
Searches done by matching the observed post-CE binaries and the modeled population in order to find 
some unique constant value of $\alpha$ might be encouraged only when they are related 
to well partitioned subclasses of post-CE binaries. 
$\lambda$ easily varies by two orders of magnitude depending on the evolutionary status 
of the donor and on the adopted core definition, and in some cases can be even negative, 
and available energy (``effective'' $\alpha$) can for some binaries exceed the orbital energy significantly.
It is unlikely therefore that any value of $\alpha$ constrained from the observations for a  group of post-CE binaries 
can either completely rule out proposed formation channels or be extrapolated to other subclasses of post-CE binaries.
It is certain that further understanding is required for such CE phases as 
the onset on CE and the slow spiral-in stage.

\bigskip

\acknowledgements N. Ivanova acknowledges support from NSERC Discovery and Canada Research Chairs programs.

\bibliography{ivanova}

\end{document}